\documentclass[12pt]{iopart}
\usepackage{iopams} 
\usepackage{epsfig}

\begin{document}


\title[Kovacs effect and FD relations]{Kovacs effect 
and fluctuation-dissipation relations in 1D kinetically 
constrained models}

\author{Arnaud Buhot\footnote[1]{Email: abuhot@cea.fr.}}

\address{UMR 5819 (CNRS, CEA, UJF), SI3M/DRFMC, 
CEA Grenoble, 17 rue des Martyrs, 38054 Grenoble 
cedex 9, France.}


\begin{abstract}
Strong and fragile glass relaxation behaviours are 
obtained simply changing the constraints of the 
kinetically constrained Ising chain from symmetric 
to purely asymmetric. We study the out--of--equilibrium 
dynamics of those two models focusing on the Kovacs 
effect and the fluctuation--dissipation relations. 
The Kovacs or memory effect, commonly observed in structural 
glasses, is present for both constraints but enhanced with 
the asymmetric ones. Most surprisingly, the related 
fluctuation-dissipation (FD) relations satisfy the FD theorem 
in both cases. This result strongly differs from the simple 
quenching procedure where the asymmetric model presents 
strong deviations from the FD theorem. 
\end{abstract}

\maketitle

\section{Introduction}

The Kovacs or memory effect has been observed by Kovacs 
himself in the 1960s on structural glassy systems~\cite{Kovacs}. 
It is a surprising memory effect of the energy (or volume) 
of the system following a particular quenching procedure. 
This quenching procedure consists in suddenly cooling a 
glassy system from a high temperature (infinite one in 
our case) to a very low intermediate temperature $T_i$ 
and letting the system relax. When the waiting time $t_w$ 
necessary for the energy of the system to reach the 
equilibrium value for a final temperature $T_f$ is 
attained ($e(t_w) = e_{eq}(T_f)$), the temperature of 
the system is set to this value $T_f \geq T_i$. 
The result found by Kovacs is that, even though the 
system is at the equilibrium energy (the volume in his 
case) corresponding to the temperature imposed, the 
system and its energy are still evolving. The system 
keeps memory of its history and of the fact that 
equilibrium is not effective. After a rapid increase 
of the energy, the system reaches the equilibrium and 
the energy decreases and levels off that equilibrium 
leading to a hump in the energy as function of time. 

In this paper, we are interested in a comparison of the 
Kovacs effect for two simple (even simplistic) models 
with respectively strong and fragile glass behaviours. 
We consider the symmetric and purely asymmetric 
kinetically constrained Ising chain (KCIC)
models~\cite{Fredrickson,Jackle}. This simple change of 
the kinetic constraints allows one to switch respectively 
from strong to fragile glass behaviour. The underlying 
equilibrium being the same for both models, a direct 
comparison of the dynamical effects is possible. 

After a short presentation of the models in section 2, 
the Kovacs effect is discussed in section 3. The effect 
is observed in both models considered but is enhanced 
in the asymmetric case. This section also contains 
simple rescaling arguments to explain this effect and 
a comparison with recent works on the Kovacs 
effect~\cite{Berthier,Berthier2,Bertin,Mossa}.
The fluctuation-dissipation (FD) relations are studied 
during the Kovacs quenching procedure and presented in 
section 4. The FD relations satisfy the FD theorem in 
both cases. These results are in strong contradiction 
with those obtained using a simple quenching procedure 
from high temperature to a low final temperature after 
a waiting time $t_w$. This last quenching procedure 
leads to FD relations satisfying the FD theorem for the 
symmetric KCIC model and for waiting times well below 
the relaxation time. In contrast, strong deviations 
from the FD theorem have been observed for the asymmetric 
KCIC model for waiting times smaller than the equilibration 
time. We give some conclusions in section 5.

\section{Presentation of the models}

In this paper, we are interested in the possible 
difference concerning the Kovacs effect due to 
strong or fragile glass behaviour. We thus consider 
the KCIC model~\cite{Fredrickson,Jackle} for which 
constraints may be chosen to model strong (for 
symmetric ones) and fragile (for a purely 
asymmetric chain) glass relaxations.

Let us consider a chain of $N$ Ising spins ($\sigma_i = 
0,1$ with $i=1,\cdots,N$) without interactions where 
spins $\sigma_i = 1$ are considered as defects. 
The corresponding Hamiltonian is thus trivial ($H = 
\sum_i \sigma_i$) as well as the equilibrium thermodynamic 
properties. The equilibrium energy at temperature $T$ or 
inverse temperature $\beta$ is given by $e_{eq}(T=1/\beta) 
= 1/(1+e^{\beta})$ which is also the concentration of 
defects or the probability to have a defect at site $i$. 
It is possible to determine exactly the probability for 
a defect to have its next defect (on the left) at a 
distance $d$
\begin{equation}
P_{eq}(d,T) = e_{eq} (1-e_{eq})^{d-1}.
\end{equation}
The first term on the right hand side of the equation 
corresponds to the probability to have a defect whereas 
the second term (with a power $d-1$) is the probability 
to have no defects in the intermediate $d-1$ sites. 
The non-interacting spins render at equilibrium the 
probabilities at each sites independent of each other 
and leads to this simple product in $P_{eq}(d,T)$. 

All these equilibrium properties are independent of any 
dynamics considered. However, the introduction of kinetic 
constraints allows one to obtain a slowing down of the 
dynamics characteristic of glassy systems before the 
equilibrium properties are reached. The probability for a 
spin to flip is constrained in the following way: in the 
symmetric case, a spin is able to flip as soon as a neighbour 
(left or right) is a defect whereas, in the asymmetric case, 
the defect has to be on the left. Such spins are also 
called spin facilitated and their probability transitions 
are given by the following equation:
\begin{equation}
P(\sigma_i \rightarrow 1 -\sigma_i) = \min(1,e^{\beta 
(2\sigma_i-1)}) (b \, \sigma_{i-1} + (1-b) \, \sigma_{i+1}). 
\label{EqPd}
\end{equation}
The first term on the right hand side corresponds to the 
usual Metropolis probability and allows one to satisfy the 
detailed balance. The second term corresponds to the general 
kinetic constraints with a probability $b$ to flip the spin 
if there is a left neighbour and $1-b$ for a right neighbour. 
The symmetric model ($b=1/2$) and the purely asymmetric 
one ($b=0$ or $1$) correspond to particular values of this 
parameter $b$. 

These models have been extensively studied (for more 
information and references on kinetically constrained models 
see the recent review by Ritort and Sollich~\cite{Ritort}). 
With the symmetric constraints, the dynamical behaviour is 
reminiscent of a strong glass with a relaxation time following 
an Arrhenius law. At sufficiently low temperature, the defects 
are mainly isolated and may be considered as simple particles 
diffusing with a temperature-dependent rate of diffusion 
$\Gamma \sim \exp(-1/T)$. The energy (or concentration of 
particles) evolves through creation and annihilation processes. 
Similar reaction-diffusion models have been introduced for a 
long time to study domain growth, coarsening and 
aging~\cite{Doering,benAvraham,Lindenberg}.
Within the asymmetric constraints, the energy barriers 
involved in the motion of defects are increasing 
logarithmically with the distance from the next 
defect~\cite{Sollich}. As a consequence, the relaxation 
time follows the B\"assler law~\cite{Bassler}: $t_{relax} 
\sim \exp(1/T^{2} \ln 2)$. Whereas the Arrhenius behaviour 
is associated to a strong glass behaviour following the 
Angell's classification~\cite{Angell}, the super-Arrhenius 
behaviour of the asymmetric model is associated to a fragile 
glass. Intermediate constraints allow the system to continuously 
crossover from fragile to strong glass behaviour~\cite{Buhot} but 
will not be considered in this study.

\section{Kovacs effect}

As already mentioned in the introduction, the Kovacs effect 
is observed following a particular quenching procedure. 
At time $t=0$, a system, equilibrated at high temperature 
($T = \infty$ in our case), is suddenly quenched to an 
intermediate low temperature $T_i$. The system starts to 
relax and the energy decreases until it reaches the 
equilibrium energy corresponding to a final temperature 
$T_f \geq T_i$ after a waiting time $t_w (T_i, T_f)$ 
defined by the following equation:
\begin{equation}
e(t_w) = \frac{1}{N}\sum_i \sigma_i(t_w) = e_{eq}(T_f) = 
(1+e^{\beta_f})^{-1}
\end{equation}
where $e(t)$ is the energy of the system at time $t$ and 
$\beta = 1/T$ is the inverse temperature. The temperature
of the system is set to $T_f$ at this waiting time $t_w$. 

If the system was characterized only by the thermodynamical 
parameters (energy, volume and temperature), we would 
expect the energy of the system to stay constant and 
equal to $e_{eq}(T_f)$ after the waiting time $t_w$. 
However, even though its energy corresponds to that 
equilibrium one at the imposed temperature, the particular 
configuration of the system at $t_w$ is still far from an 
equilibrium configuration at $T_f$. As a consequence, the 
energy is still evolving after $t_w$.

\subsection{The symmetric KCIC model}

\begin{figure}[t]
\begin{center}
\epsfig{file=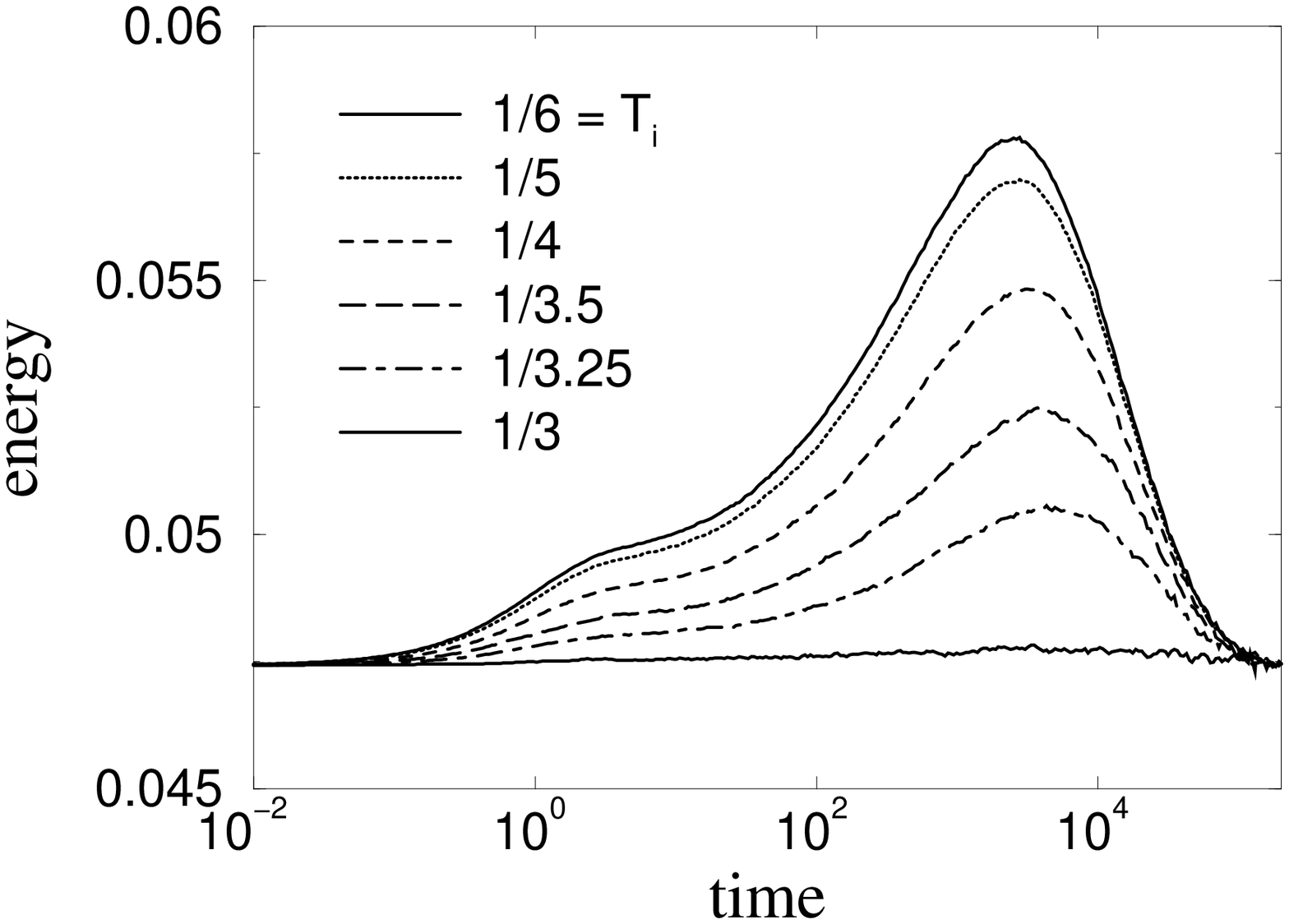, width=6.9cm} 
\epsfig{file=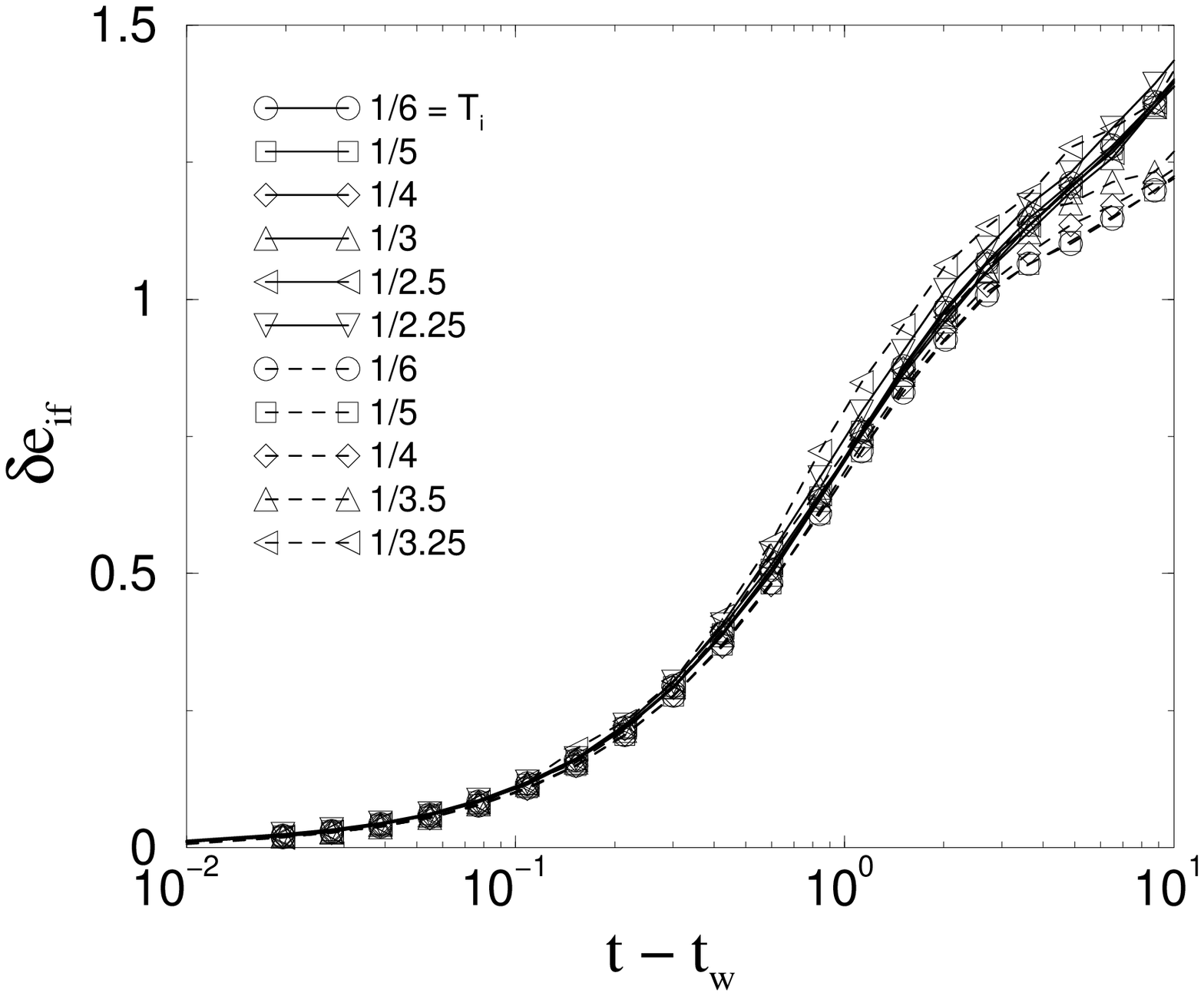, width=6cm}
\caption{Left : Kovacs effect for a final temperature 
$T_f = 1/3$ and different intermediate temperatures 
$T_i$ within the symmetric KCIC model: the energy as 
function of time elapsed since the waiting time is 
plotted. Right: the rescaled energy $\delta e_{if}$ 
for short times (see text for definition) is plotted 
for two final temperatures $T_f = 1/2$ (full lines) 
and $1/3$ (dashed lines) and for different intermediate 
temperatures. \label{KovacsSym}}
\end{center}
\end{figure}

The figure~\ref{KovacsSym} illustrates the Kovacs effect 
for the symmetric KCIC model~\cite{Simul}. As can be seen 
on the left panel, the Kovacs effect is an increasing 
function of the difference between the intermediate and 
final temperatures. A two timescale behaviour is also 
clearly observed. The energy increases on a timescale $t
\sim O(1)$ to reach a plateau followed by a further increase 
up to a maximum. The energy finally decreases to reach the 
equilibrium energy at temperature $T_f$. Note also that 
the maximum in energy decreases and shifts to slightly higher 
times for higher intermediate temperatures $T_i$. This result 
was already observed for the volume by Kovacs~\cite{Kovacs} 
and for the energy in recent simulations~\cite{Berthier,Berthier2}.

The maximum of the energy is obtained for a timescale 
$t_m \sim O(\exp (3/T_f))$. This timescale is of 
the same order as the equilibrium timescale $t_{eq}$ 
(time to reach the equilibrium after a rapid quench 
from a high temperature to a low temperature 
$T_f$)~\cite{Jackle,Ritort}. This $\exp(3/T_f)$ 
dependence of $t_{eq}$ is explained considering that 
the energy $e(t)$ after such a quench decays like the 
annihilation process $A + A \rightarrow A$ in one 
dimension with a diffusion rate $\Gamma$: $e(t) \sim 
(\Gamma t)^{-1/2}$. The energy thus reaches that equilibrium 
($e_{eq}(T_f) \sim e^{-1/T_f}$) after an equilibrium 
time $t_{eq} \sim \Gamma^{-1} \, e_{eq}^{-2} \sim e^{3/T_f}$ 
due to the temperature-dependent diffusion rate ($\Gamma 
\sim e^{-1/T_f}$). With a similar argument it is possible 
to estimate the waiting time $t_w(T_i,T_f) \sim \Gamma^{-1}
(T_i) \, e_{eq}^{-2}(T_f) \sim e^{1/T_i} \, e^{2/T_f}$ where 
the equilibrium energy to reach is $e_{eq}(T_f) \sim e^{-1/T_f}$ 
but the diffusion rate is $\Gamma(T_i) \sim e^{-1/T_i}$ since 
the temperature of the system before $t_w$ is set to $T_i$. 
We recover the equilibrium time when $T_i = T_f$. The estimate 
of the waiting time agrees qualitatively with the values
obtained from numerical simulations for the different couples
of intermediate and final temperatures considered.

Let us now analyse the short timescales behaviour. 
The fast increase of the energy is related to the 
probability $p_i$ to have neighbouring defects. 
Such a probability is given at $t_w$ by $p_i = 
e(t_w) \, e_{eq}(T_i)$ where $e(t_w) = e_{eq}(T_f)$ 
is the probability to have a defect at $t_w$ and 
$e_{eq}(T_i)$ is the probability for its neighbour 
to also be a defect. This last term corresponds to 
the equilibrium probability $P_{eq}(d=1,T_i)$ due 
to the fact that the timescale to equilibrate the 
concentration of neighbouring defects (or non-constrained 
defects) is $t \sim O(1) \ll t_w$ and has thus already 
equilibrated at $t_w$. When the temperature is increased 
from the low temperature $T_i$ to a higher one $T_f$, 
on a similar timescale $t - t_w \sim O(1)$ independent 
of the temperature, the probability to have neighbouring 
defects reaches its equilibrium value $p_f = e_{eq}^2(T_f)$. 
Furthermore, the timescale for the motion of defects is 
$\Gamma^{-1}(T_f) \sim \exp(1/T_f) \gg 1$. As a consequence, 
the change of probability from $p_i$ to $p_f$ is mainly 
obtained from the creation of defects and the energy as 
function of time may be expressed as
\begin{equation}
e(t-t_w) = e_{eq} + (p_f - p_i) \delta e_{if}(t-t_w)
\end{equation}
where we expect $\delta e_{if} (t-t_w)$ to be independent 
of both intermediate and final temperatures for $t-t_w 
\sim O(1)$ and to start to differ at $\delta e_{if} 
\simeq 1$ for different $T_f$. This prediction is verified 
on the right panel of figure~\ref{KovacsSym} where 
$\delta e_{if}$ is plotted for different couples $(T_i, T_f)$
on short times $t-t_w$. 

\begin{figure}[t]
\begin{center}
\epsfig{file=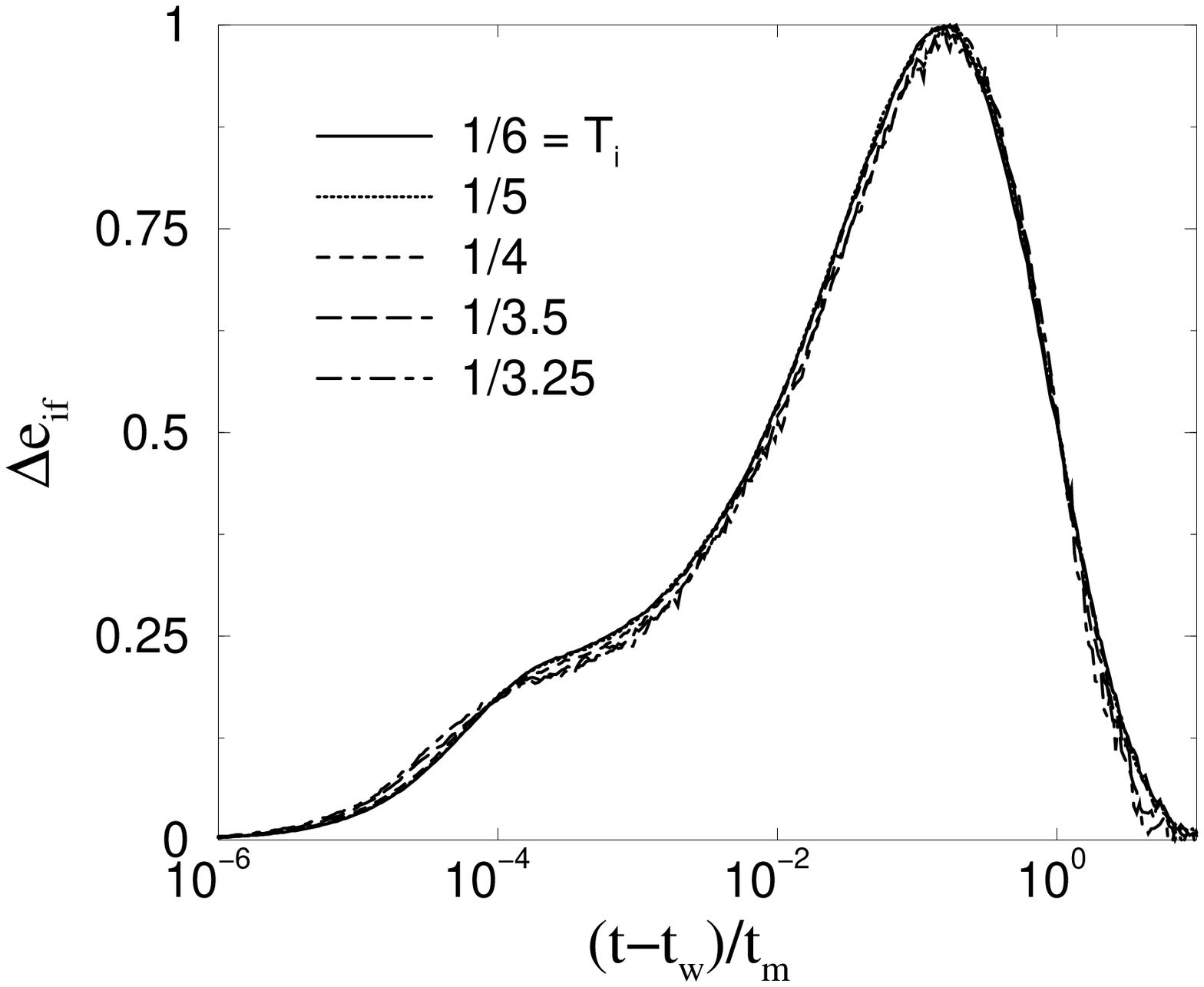, width=6.4cm}
\epsfig{file=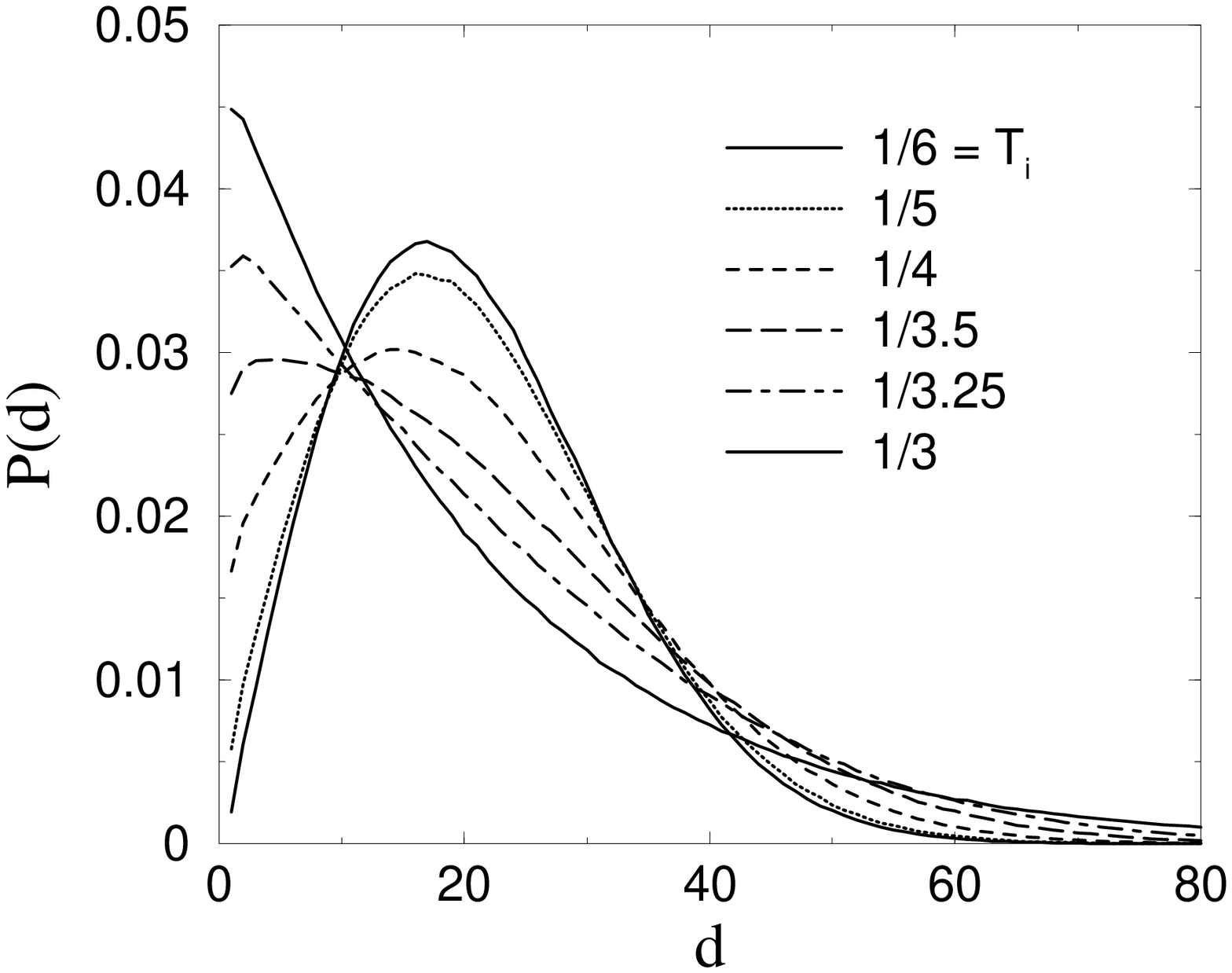, width=6.4cm}
\caption{Left : Rescaled energy $\Delta e_{if}$ as function 
of the rescaled time $(t-t_w)/t_m$ (see text for definition) 
and the distribution of distances between defects
at the different waiting times $t_w(T_i, T_f)$. For both 
figures, the model considered is the symmetric KCIC and 
the final temperature is $T_f = 1/3$. \label{KovacsSym2}}
\end{center}
\end{figure}

On longer timescales, the activated regime is involved. 
The defects may then be considered as simple particles 
moving with a temperature-dependent diffusion rate 
$\Gamma_f \sim \exp(-1/T_f)$. The number of particles
(or energy) evolves through creation of neighbouring 
defects and annihilation when two defects collide ($A+A 
\leftrightarrow A$). As we have seen with the short 
timescales behaviour, the energy increases after the 
waiting time $t_w$. This increase is not restricted 
to the short timescales but continues on longer 
timescales and is due to the fact that the distribution 
of distances between defects $P(d)$ at the different 
waiting times $t_w(T_i, T_f)$ is far from the equilibrium 
one $P_{eq} (d,T_f)$ (see Eq.(\ref{EqPd}) for definition). 
This difference is evident on the right panel of 
figure~\ref{KovacsSym2}. The distribution $P(d)$ 
converges to that equilibrium when the intermediate 
temperature reaches the final one as it should. 
One important point is that the whole range of the 
distribution is affected. Defects with neighbouring
defects at all distances relax with the same typical 
timescale due to the single energy barrier involved 
(the motion of a defect from one site to the next one 
is independent of the distance from its neighbouring
defect). 

The interpretation with diffusive particles and with 
annihilation-creation processes is similar to coarsening
models. The only differences with the 1D Ising model
concern the diffusive rate which is temperature-dependent
in our case and the creation-annihilation processes which 
are of the type $A + A \leftrightarrow \emptyset$ in the
Ising model. This analogy to domain growth models allows 
us to use the same rescaling for long timescales of the 
Kovacs effect. In recent works~\cite{Berthier,Berthier2,Bertin}, 
it has been shown that, for domain growth models, the energy 
shift from the equilibrium energy may be rescaled for all 
intermediate temperatures $T_i$ in the following way:
\begin{equation}
e(t) = e_{eq}(T_f) + e_m \ \Delta e_{if}
\left(\frac{t-t_w}{t_m}\right) \label{rescal}
\end{equation}
where $t_m$ is the time for which the shift in energy from the 
equilibrium energy already diminished by a factor two from its 
maximum value
\begin{equation}
e(t_m) = e_{eq}(T_f) + e_m/2.
\end{equation}
As can be seen on the left panel of figure~\ref{KovacsSym2}, 
this rescaling is correct for the long timescales. For short 
timescales, a small deviation from this rescaling is observed. 

\subsection{The asymmetric KCIC model and comparison}

\begin{figure}[t]
\begin{center}
\epsfig{file=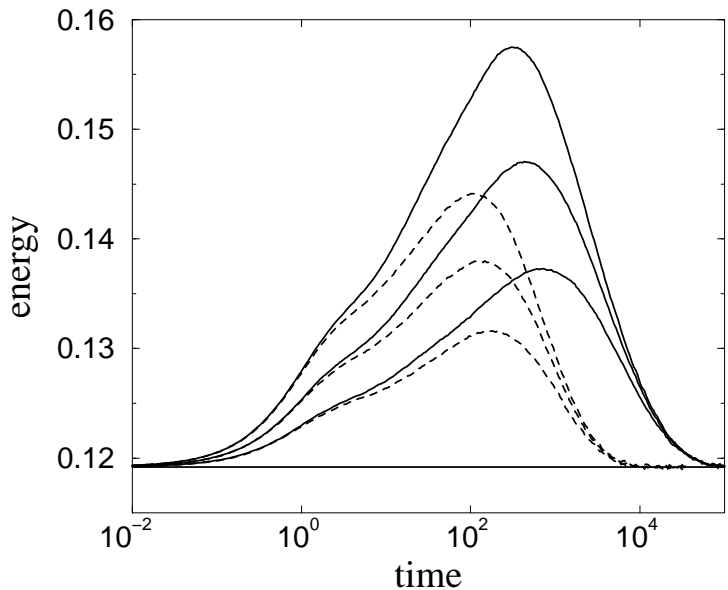, width=10cm} 
\caption{Comparison of the Kovacs effect for the symmetric 
(dashed lines) and the asymmetric (full lines) KCIC models. 
The energy as function of time elapsed since the waiting 
time is plotted for a final temperature $T_f = 1/2$ and 
different intermediate temperatures $T_i = 1/4, 1/3$ and 
$1/2.5$.\label{Compar}}
\end{center}
\end{figure}

\begin{figure}[t]
\begin{center}
\epsfig{file=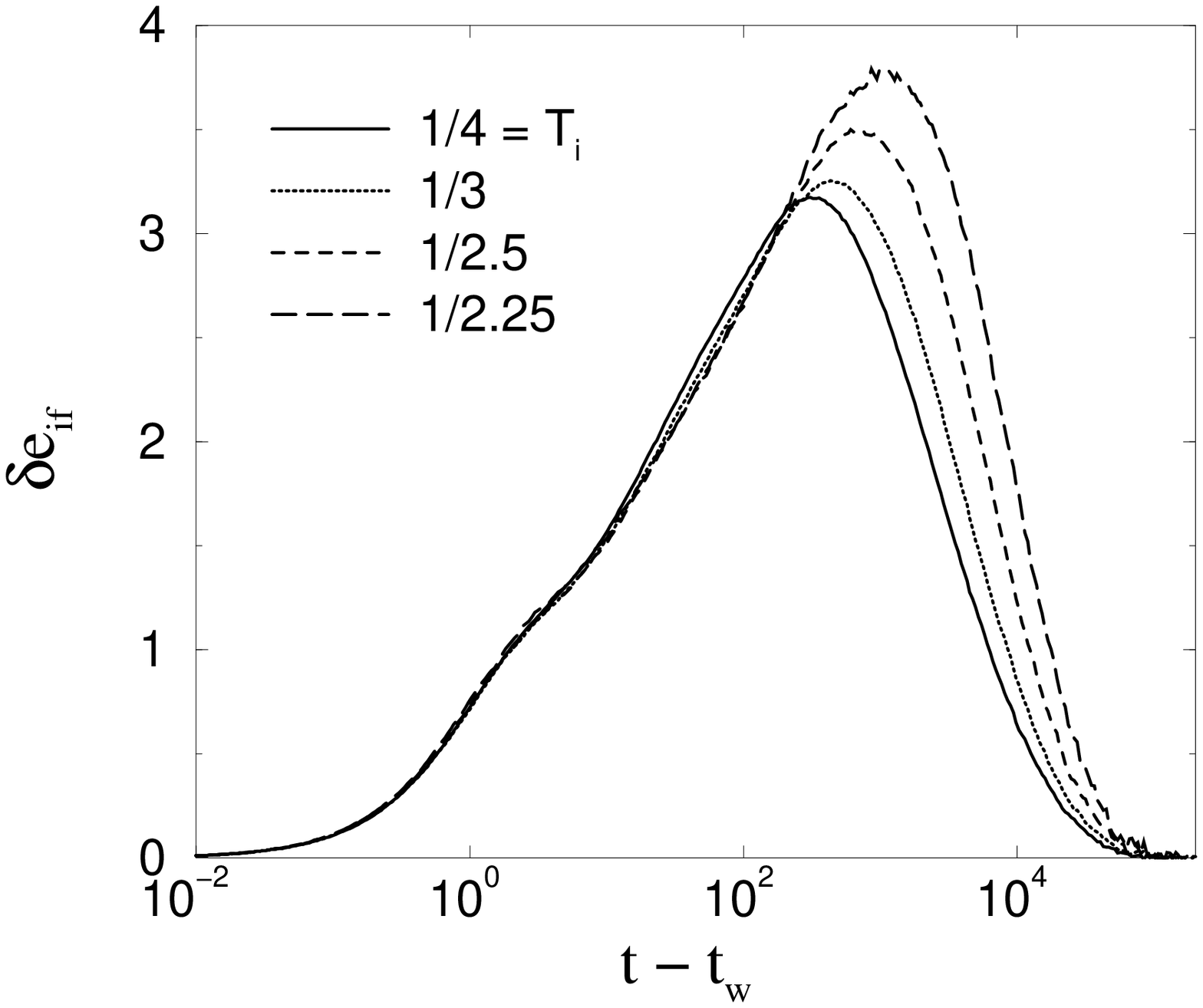, width=6.4cm} 
\epsfig{file=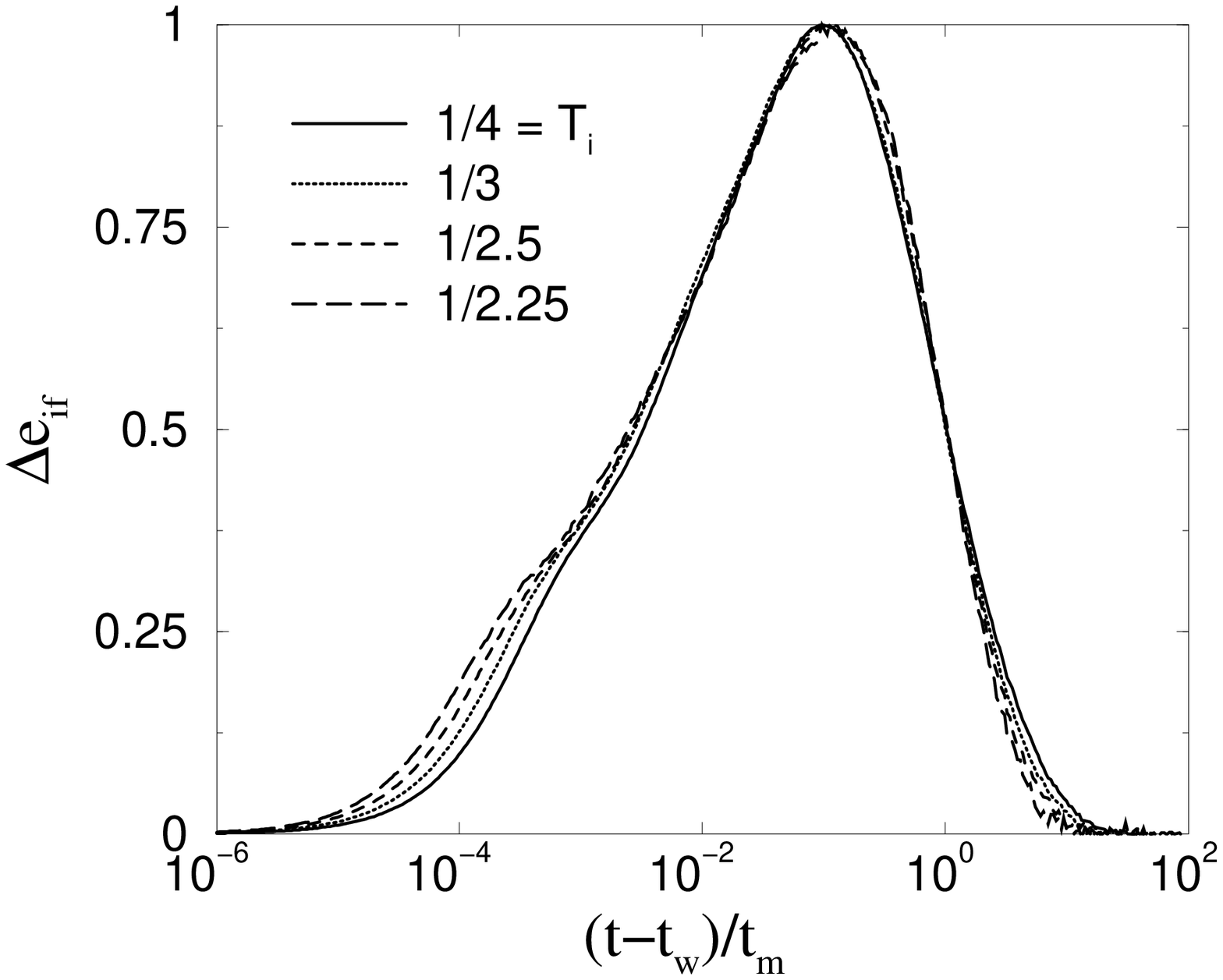, width=6.4cm}
\caption{Kovacs effect for the final temperature $T_f = 
1/2$ and different intermediate temperatures $T_i$ with 
the asymmetric KCIC model. On the left figure are plotted 
the rescaled energy $\delta e_{if}$ for short timescales as 
function of time since the waiting time and, on the right 
figure, the rescaled energy for long timescales as function 
of the rescaled time. \label{KovacsAsym}}
\end{center}
\end{figure}

In figure~\ref{Compar}, we compare the Kovacs effects for 
the symmetric and asymmetric KCIC models. The energy as 
function of the time $t-t_w$ since the final temperature 
$T_f = 1/2$ was set to the system is plotted for different 
intermediate temperatures $T_i = 1/4, 1/3$ and $1/2.5$. 
As can be seen, the short timescales behaviour (non-activated)
is similar in both models. The same rescaling introduced for 
the symmetric KCIC model works perfectly on short timescales 
for the asymmetric KCIC model (see left panel of 
figure~\ref{KovacsAsym}). The argument developed to obtain
the rescaling is independent of the dynamical constraints as 
soon as the motion of particles occurs on timescales larger 
than the spontaneous creation of neighbouring defects. 

On longer timescales, when activation plays an important
role, the Kovacs effect starts to differ between the 
symmetric and asymmetric KCIC models. The main result 
is a higher maximum of the energy for the asymmetric case. 
The timescales involved are also not surprisingly different 
due to equilibration times which strongly differ from strong 
to fragile glass behaviour. This can also be traced out on the 
values of the waiting times $t_w(T_i, T_f)$ larger for the 
asymmetric KCIC model (fragile glass behaviour) than for the 
symmetric KCIC model (strong glass behaviour). From the 
analysis of Sollich and Evans~\cite{Sollich}, it is 
possible to have an estimate of the waiting time 
$t_w(T_i,T_f)$ for the asymmetric KCIC model. The same 
creation-annihilation processes occur as for the symmetric 
KCIC model but the motion of particles may not be considered 
as simple diffusion due to the increasing energy barriers 
with the distance between defects. An anomalous coarsening 
occurs where the energy decreases like $e(t) \sim 
t^{-T_i \ln 2}$ with a temperature-dependent exponent 
in contrast to normal coarsening. The waiting time
is obtained when the energy reaches the equilibrium 
energy $e_{eq}(T_f) \sim e^{-1/T_f}$ leading to 
$t_w(T_i,T_f) \sim \exp (1/T_i T_f \ln 2)$.
Note that once again we recover the equilibrium timescale
$t_{eq}(T_f) \sim \exp (1/T_f^2 \ln 2)$ when $T_i = T_f$.
The estimate for the waiting time agrees qualitatively 
with the values obtained from the numerical simulations
for different couples of intermediate and final temperatures.

We use the rescaling for the long timescales of coarsening 
models (\ref{rescal}) (see right panel of figure~\ref{KovacsAsym}) 
as for the symmetric KCIC model. We observe some deviations 
from a perfect rescaling even for the long timescales. This 
difference is due to the  anomalous coarsening behaviour of 
the asymmetric KCIC model. The failure of the rescaling on 
short timescales is also more evident. 

\section{Fluctuation-dissipation relations}

\begin{figure}[t]
\begin{center}
\epsfig{file=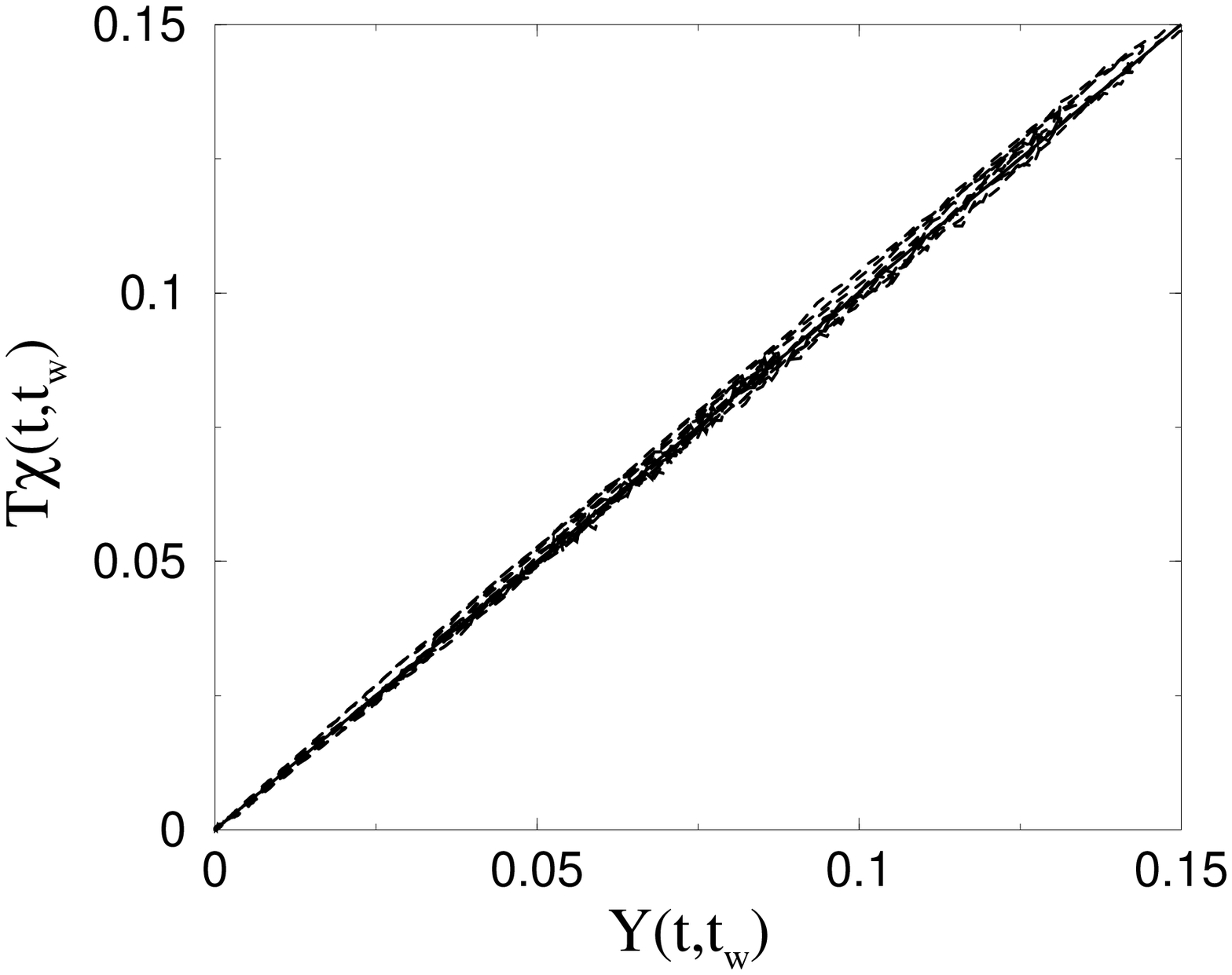, width=6.5cm}
\epsfig{file=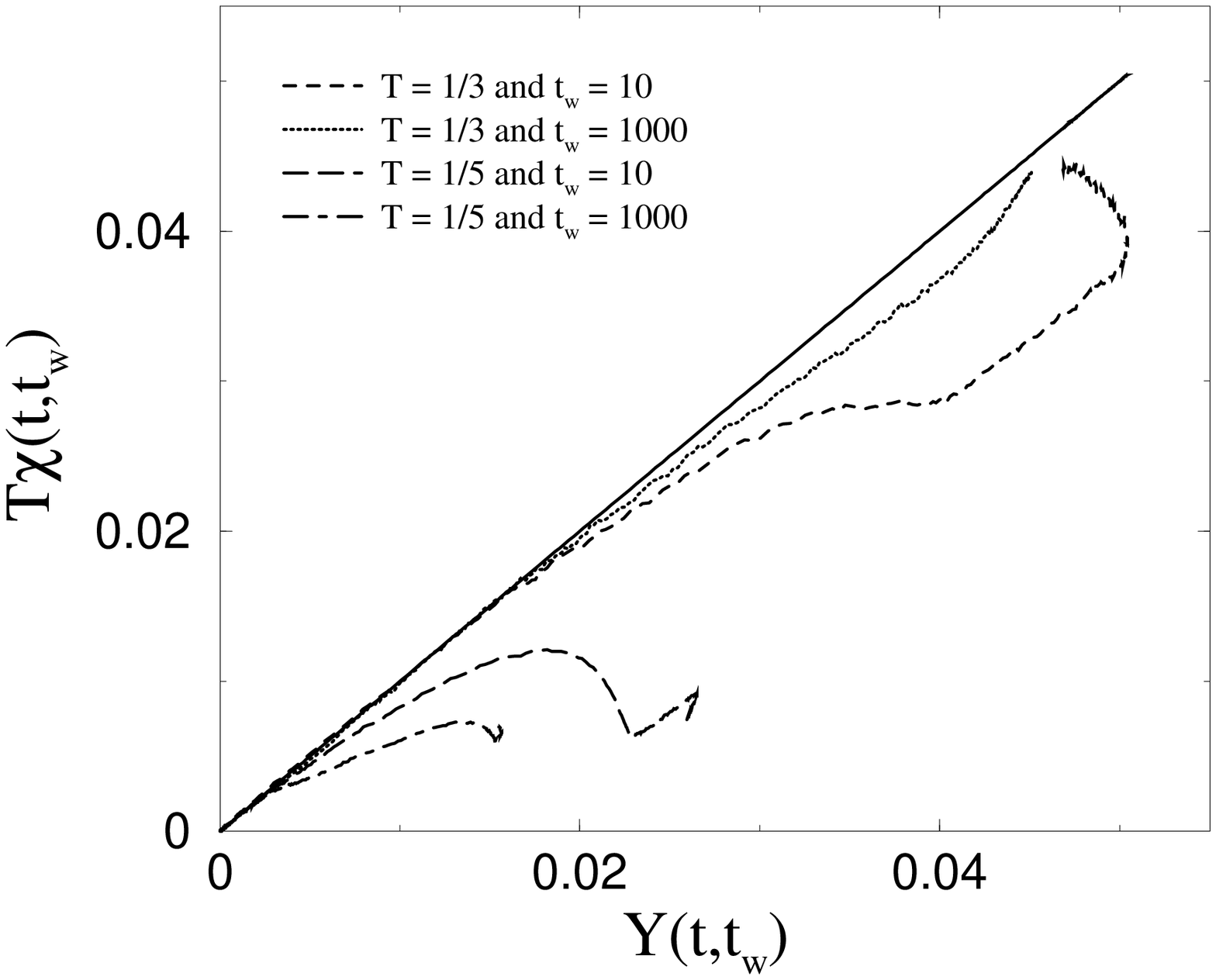, width=6.3cm}
\caption{Fluctuation-dissipation relations for the symmetric 
(left) and asymmetric (right) KCIC models with a usual 
quenching procedure (see text for explanation). A set of 
final temperatures ($T_f=1/\beta_f = 1/3$ and $1/6$) and 
waiting times ($t_w = 10, 100, 1000$ and $10000$) have been 
studied and all couples of temperatures-waiting times 
(dotted lines) satisfy the FD theorem (full line) in the 
symmetric case. A set of final temperatures ($T_f=1/\beta_f 
= 3$ and $5$) and waiting times ($t_w = 10$ and $1000$) have 
been studied and strong deviation from the FD theorem (full 
line) have been observed for all couples of temperatures-waiting 
times in the asymmetric case. \label{FDT1}}
\end{center}
\end{figure}

The violation of the fluctuation-dissipation (FD) theorem
is usually present in glassy systems due to the 
out--of--equilibrium dynamics (see for example the recent 
review by Crisanti and Ritort~\cite{Crisanti2} and the 
references therein). FD relations have already been considered 
for the symmetric and asymmetric KCIC models in~\cite{Crisanti} 
with the usual quenching procedure. A more careful analysis of 
the symmetric case leads to the conclusion that even far from 
equilibrium (at least after a short transient), the FD 
relations satisfy the FD theorem for all temperatures and 
waiting times~\cite{Buhot3}. This result was explained by 
the fact that defects are mainly isolated at low temperatures 
after a short transient and behave as in local equilibrium 
(with a temperature-dependent diffusive rate $\Gamma \sim 
\exp(-1/T)$) even if the number of defects is larger than 
in equilibrium. This argument is no longer valid for the 
asymmetric case where the motion of a defect is dependent 
on the position of its left neighbour defect and is thus 
sensitive to the total number of defects or at least 
to the distribution of distances between defects which 
is out--of--equilibrium. The differences between the FD 
relations for both models are evident in figure~\ref{FDT1}
following the usual quenching procedure. We consider such 
FD relations after the Kovacs quenching procedure to check 
if this discrepancy between the two models is also present.

Let us first explain how those FD relations are determined.
The integrated response is determined following the now standard
procedure introduced by Barrat~\cite{Barrat}. A pertubation
$\delta H(t) = - h(t)\, \sum_i \varepsilon_i \sigma_i$ is added to 
the Hamiltonian where $h(t) = h \, \Theta (t - t_w)$ is a field of 
strength $h$ introduced after the waiting time $t_w$ ($\Theta (x)$ 
is the Heaviside function). Random fields are set to each site 
through the random variables $\varepsilon_i = \pm 1$ with identical 
probabilities. The integrated response is then
\begin{equation}
\chi(t,t_w) = \frac{1}{h N} \sum_i \overline{\varepsilon_i \langle 
\sigma_i (t) \rangle_h}
\end{equation}
with the overline standing for an average over the random variables 
$\varepsilon_i$ and the brackets for the dynamical average in the 
presence of the perturbation. Note that in these kinetically 
constrained models different dynamics may be considered after 
the introduction of the perturbation. We considered the modified 
Metropolis one discussed in~\cite{Buhot4} which gives similar 
results to the Metropolis one in this case~\cite{Simul}.

The integrated response has to be compared to the corresponding 
correlations to check the validity of the FD theorem. In the case 
where the energy is still evolving the correct correlations to 
consider are the connected ones
\begin{equation}
C(t,t_w) = \frac{1}{N} \sum_i \langle \sigma_i(t) \sigma_i(t_w) 
\rangle - e(t) e(t_w)
\end{equation}
where $e(t)$ is the energy at time $t$ and the brackets stand 
for the dynamical average without the perturbation. 
In equilibrium, the integrated response and correlations only 
depend on the difference of times and satisfy: $T \chi(t-t_w) 
= C(0)-C(t-t_w)$. It has been shown in~\cite{Buhot3} that a 
correct generalization of this expression for systems 
out--of--equilibrium is
\begin{equation}
T \chi(t,t_w) = f(Y(t,t_w))
\end{equation}
with $Y(t,t_w) \equiv C(t,t) - C(t,t_w)$. If the FD theorem 
is satisfied the function $f(x) = x$. Thus, a parametric plot 
of the integrated response with respect to the function $Y$ 
allows one to check the validity of the FD theorem.
Note that in these kinetically constrained systems, 
$C(t,t) = e(t)-e^2(t)$ depends on time. This explains the 
parametric plot with respect to $Y(t,t_w)$ instead of the 
two-time correlation $C(t,t_w)$. A slope $X$ different than 
$1$ in this parametric plot would suggest the existence of 
an effective temperature $T_{eff} = T/X$.

\begin{figure}[t]
\begin{center}
\epsfig{file=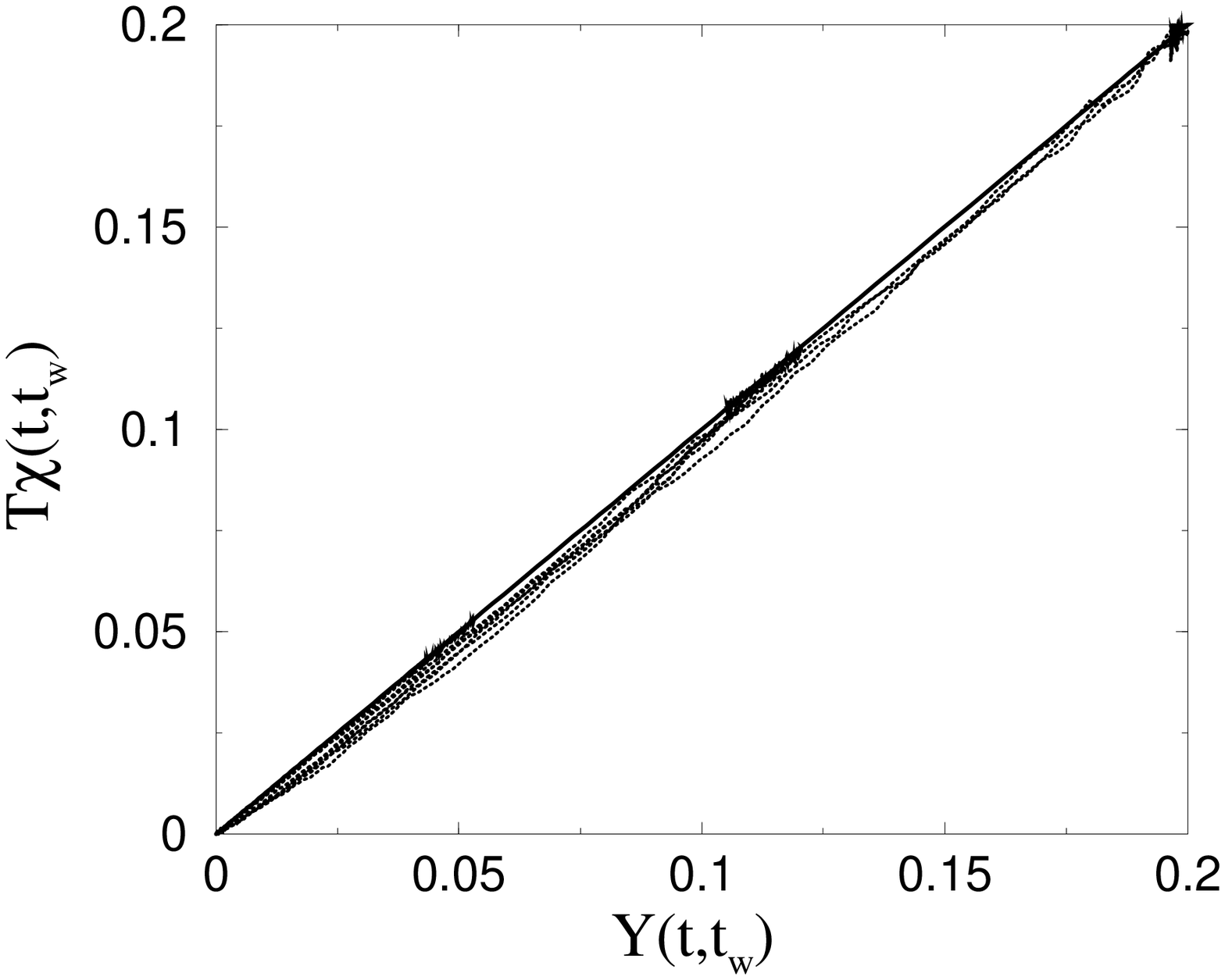, width=6.4cm}
\epsfig{file=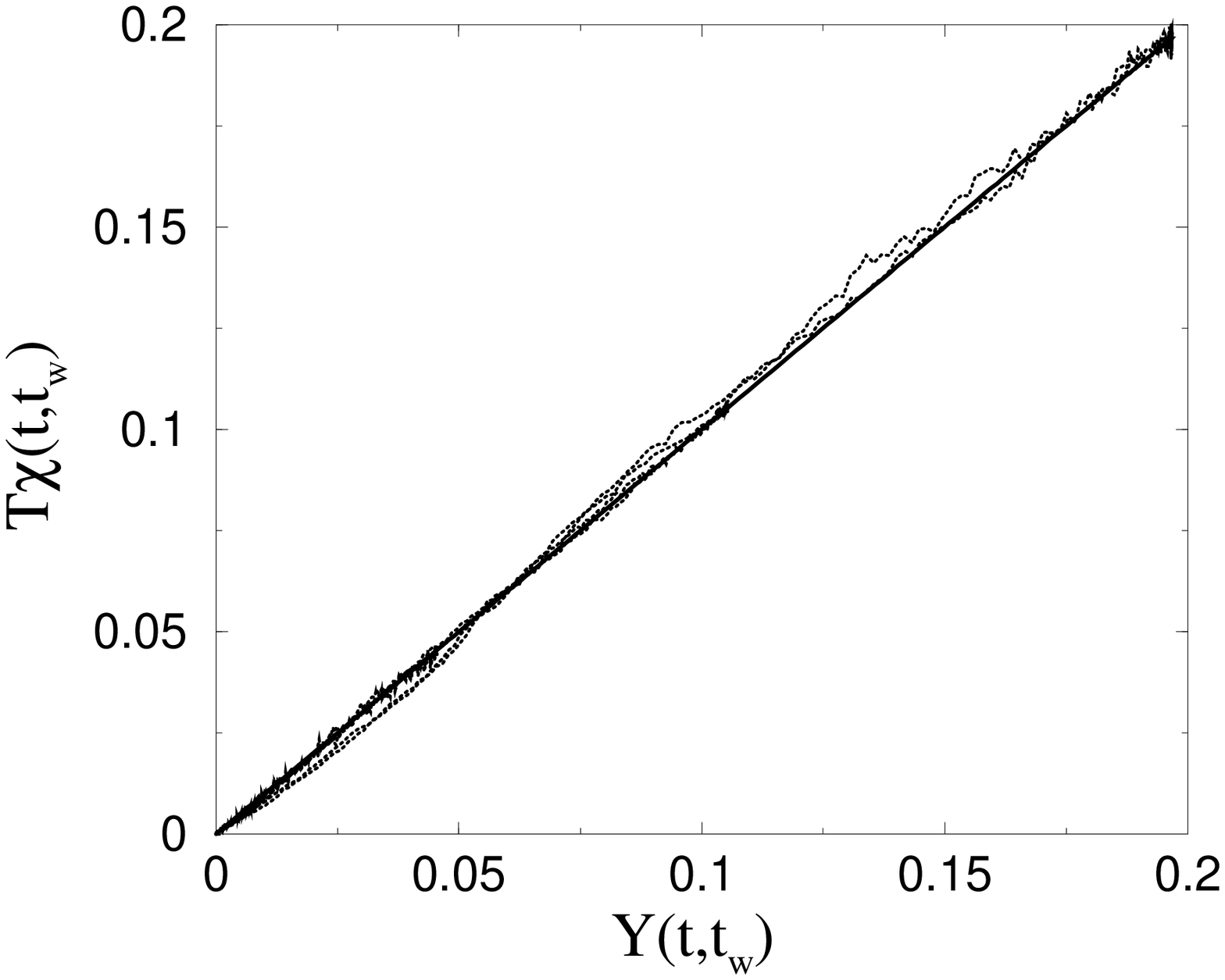, width=6.4cm}
\caption{Fluctuation-dissipation relation for the symmetric 
(left) and asymmetric (right) KCIC models with a Kovacs 
quenching procedure (see text for explanation). A set of 
intermediate and final temperatures ($T_i=1/\beta_i$, 
$T_f=1/\beta_f$) have been studied and all couples of 
temperatures (dotted lines) satisfy the FD theorem 
(full line). Final and intermediate inverse temperatures 
considered for the symmetric case: $\beta_f = 1,2$ and 
$3$ and $\beta_i = 3,4$ and $5$ and for the asymmetric 
case: $\beta_f = 1,2$ and $3$ and $\beta_i = 3$ and $4$.
\label{FDT}}
\end{center}
\end{figure}

In figure~\ref{FDT1}, we have plotted the FD relations 
for both symmetric (left) and asymmetric (right) KCIC 
models following the usual quenching procedure. 
At $t=0$, the system is quenched from an infinite temperature 
to a low temperature $T_f$. The perturbation is introduced 
at a waiting time $t_w$ after this quench. On the left 
panel, the FD relations for the symmetric KCIC model 
satisfy the FD theorem for a large set of temperatures 
and waiting times (as soon as $t_w > 1$). The validity 
of the FD theorem is far from obvious if we keep in mind 
that both the integrated response and $Y(t,t_w)$ are 
non--monotonic. However, they compensate each other to 
follow the FD theorem curve at all times. 
The non--monotonic behaviour of the response is due to 
the decrease of the energy on long timescales. 
When rescaled by the energy, the response is continously
increasing.

For the asymmetric KCIC model, the FD relations show strong 
deviation from the FD theorem. The non-monotonic integrated 
response cannot be accounted for by the difference of 
correlations $Y(t,t_w)$ as it is for the symmetric KCIC model. 
It leads to different humps in the FD relations corresponding 
to the different plateaus observed in the decay of the energy.
The increasing energy barriers ($\Delta E  = 1, 2, ...$) lead 
to those different plateaus in the energy decay after a quench 
in temperature~\cite{Sollich,Buhot}. The FD theorem is recovered 
for sufficiently large waiting times (for $t_w > 10^4$ when 
$T_f = 1/3$ for example).

\begin{figure}[t]
\begin{center}
\epsfig{file=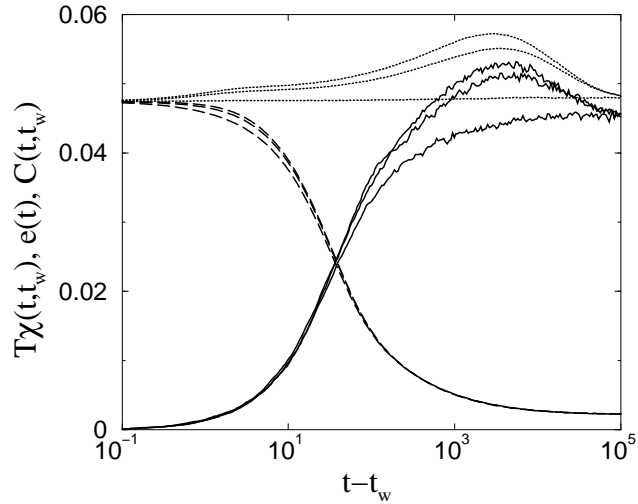, width=8.5cm}
\caption{Responses (full lines), energies (dotted lines) 
and correlations (dashed lines) for different intermediate 
temperatures $T_i$ (from top to bottom $T_i = 1/5, 1/4$ 
and $1/3$) for the symmetric KCIC model with $T_f = 1/3$. 
\label{resp}}
\end{center}
\end{figure}

Let us now consider the FD relations for the symmetric 
(left) and asymmetric (right) KCIC models after the 
Kovacs quenching procedure (see figure~\ref{FDT}). 
As can be seen in both cases the FD theorem is verified. 
This is in strong contrast with the evolution of the 
energy which shows a deviation from equilibrium.
This deviation is sufficiently important to exclude the 
fact that it does not show up on the FD relations. 
In figure~\ref{resp} we can observe that the responses 
for different intermediate temperatures as well as the 
energies and the correlations are different. However, 
the FD theorem is satisfied. Note for example that 
when $T_i$ is highly slower than $T_f$, the response 
shows a hump which may be related directly to the fact 
that the energy is not constant. An increasing energy, 
corresponding to a larger number of defects, tends to 
increase the response. The further decrease of the response 
is then related to the same decrease in the energy. 
The consequence on the FD relations is that the parametric 
curve increases from the origin to a maximum before a 
small decrease occurs but always following the line 
$f(x) = x$ corresponding to the FD theorem curve.

The fact that the FD relations follow the FD theorem curve 
for the symmetric KCIC model is not surprising since it is 
already the case after a usual quenching procedure. A simple 
explanation suggesting that defects are in local equilibrium 
has been given. The surprise comes from the asymmetric KCIC 
model where the FD relations differ strongly from the FD 
theorem curve for most waiting times with the usual quenching 
procedure. With the Kovacs one, the FD relations follow the FD 
theorem curve. A possible argument would be the following. 
The waiting times involved in the Kovacs quenching procedure 
are sufficiently large that the out--of--equilibrium 
configuration may be considered as equivalent to a simple 
fluctuation in the system at equilibrium. In other words the 
linear regime is attained and the FD theorem is satisfied. 
However, this argument would not explain the shift in 
energy observed. 

\section{Conclusion}

In this paper, we have studied the Kovacs effect for the symmetric 
and asymmetric KCIC models with respectively strong and fragile 
glass behaviours. The effect is stronger in the asymmetric case.
The symmetric KCIC model satisfies the same rescaling as domain 
growth models for long timescales. Some small deviations are present 
for the asymmetric KCIC model and may be explained due to the 
anomalous coarsening. The short timescales behaviour is identical 
for both models and a simple argument allows one to rescale all 
simulations for different intermediate and final temperatures 
on the same leading curve which happens to be independent of 
the dynamics (identical for the symmetric and asymmetric KCIC 
models). 

We have also studied the FD relations for both models after a 
Kovacs quenching and a usual one. After the Kovacs quenching, 
the related FD relations verify surprisingly the FD theorem 
for both models in strong contrast with the results for the 
asymmetric KCIC model using the usual quenching procedure. 
No clear explanation has been found for this behaviour. 
It could be interesting to have a look at the Kovacs effect 
and the corresponding FD relations using the models introduced 
in~\cite{Buhot4} where analytical results may be expected due 
to the possible exact mean-field solution.

\section*{Acknowledgments}
The author would like to thank R. Calemczuk for useful discussions
and a careful reading of the manuscript.


\begin{thebibliography}{99}
\bibitem{Simul}
Numerical simulations have been done using the continuous time 
algorithm~\cite{BKL} on systems of size $N = 10^6$ and averaged 
on $20-50$ runs. For the FD relations, the strength of the field
$h=0.05$ was sufficient to be in the linear regime.
\end{thebibliography}
\end{document}